\documentclass[epj]{svjour}

\usepackage{graphicx}
\usepackage{fancyhdr}
\def\makeheadbox{{
 }}
\def\mytitle{My title}
\def\myauthors{My name}
\def\mytype{My type of session}
\def\mysession{My session}

\def\mytitle{Search for Supersymmetric Neutral Higgs Bosons at the Tevatron} 
\def\myauthors{Tim Scanlon}    
\def\mytype{Contributed Talk}
\def\mysession{Colliders - Higgs Phenomenology}

\begin{document}
\title{Search for Supersymmetric Neutral Higgs Bosons at the Tevatron}
\author{Tim Scanlon
\thanks{\emph{Email: tim.scanlon@imperial.ac.uk}}%
\thanks{\emph{On behalf of the CDF and D\O\ Collaborations.}}
}                     
%
%
\institute{Imperial College London, Blackett Laboratory, Physics
Department, Prince Consort Road, London SW7 2AZ, United Kingdom.}
%
\date{}
\abstract{ Recent preliminary results obtained by the CDF and D\O\
Collaborations on searches for Higgs bosons beyond the Standard
Model at Run II of the Tevatron are discussed. The data,
corresponding to integrated luminosities of up to $1$ fb$^{-1}$, are
compared to theoretical expectations. No significant excess of
signal above the expected background is observed in any of the
various final states examined, and so limits at $95$\% Confidence
Level (CL) are presented.
\PACS{
      {PACS-key}{discribing text of that key}   \and
      {PACS-key}{discribing text of that key}
     } 
} 
\maketitle
%

\section{Introduction}
\label{intro}

The search for Higgs bosons is one of the main challenges for
particle physics and as such a high priority for the upgraded CDF
and D\O\ detectors at Run II of the Tevatron. Higgs boson production
cross sections in the Standard Model (SM) are small at the Tevatron.
However many models beyond the SM, including Supersymmetry, predict
larger Higgs production cross sections, some within reach even with
the present data sets.

The Minimal Supersymmetric extension of the SM (MSSM) \cite{mssm}
introduces two Higgs doublets and so contains five physical Higgs
bosons. Two of them are CP-even scalars, $h$ and $H$, of which $h$
is the lighter and SM like. The other three consist of a charged
Higgs pair, $H^{\pm}$, and a CP-odd scalar, $A$, the mass of which
($m_{A}$) is one of the two free parameters of the model at tree
level. The production cross section of the Higgs in the MSSM is
proportional to the square of the second free parameter of the
model, tan$\beta$, the ratio of the two vacuum expectation values of
the Higgs doublets. Large values of tan$\beta$ thus result in
significantly increased production cross sections compared to the
SM.  Moreover, in the large tan $\beta$ limit one of the CP-even
scalars and the CP-odd scalar are degenerate in mass, leading to a
further cross section enhancement. The main production mechanisms
for such neutral Higgs bosons are the $gg, b\bar{b} \rightarrow
\phi$ and $gg, q\bar{q} \rightarrow \phi + b\bar{b}$ processes,
where $\phi = h, H, A$. The branching ratio of $\phi \rightarrow
b\bar{b}$ is around 90\% and $\phi \rightarrow \tau^{+}\tau^{-}$ is
around 10\%. The overall experimental sensitivity is however similar
for the two channels, due to the lower background in the $\tau$
channel.

Other extensions to the SM such as Top-color \cite{topcolor} or
Fermiophobic Higgs models \cite{haber} also lead to enhanced decays
of Higgs $\rightarrow \gamma\gamma$ which is negligible in the SM.

There are consequently a number of non-SM Higgs searches already
being actively pursued with the first fb$^{-1}$ of data collected
during Run II. This note summarises these analyses. All results are
preliminary and more information is available from the public pages
of CDF and D\O\ \cite{cdf_web_page,dzero_web_page}.

\section{Limits on neutral SUSY Higgs at high tan$\beta$}
\label{sec:1}

\subsection{Higgs $\rightarrow \tau^{+}\tau^{-}$}
\label{sec:2}

\begin{figure}
\begin{center}
\includegraphics[width=0.48\textwidth]{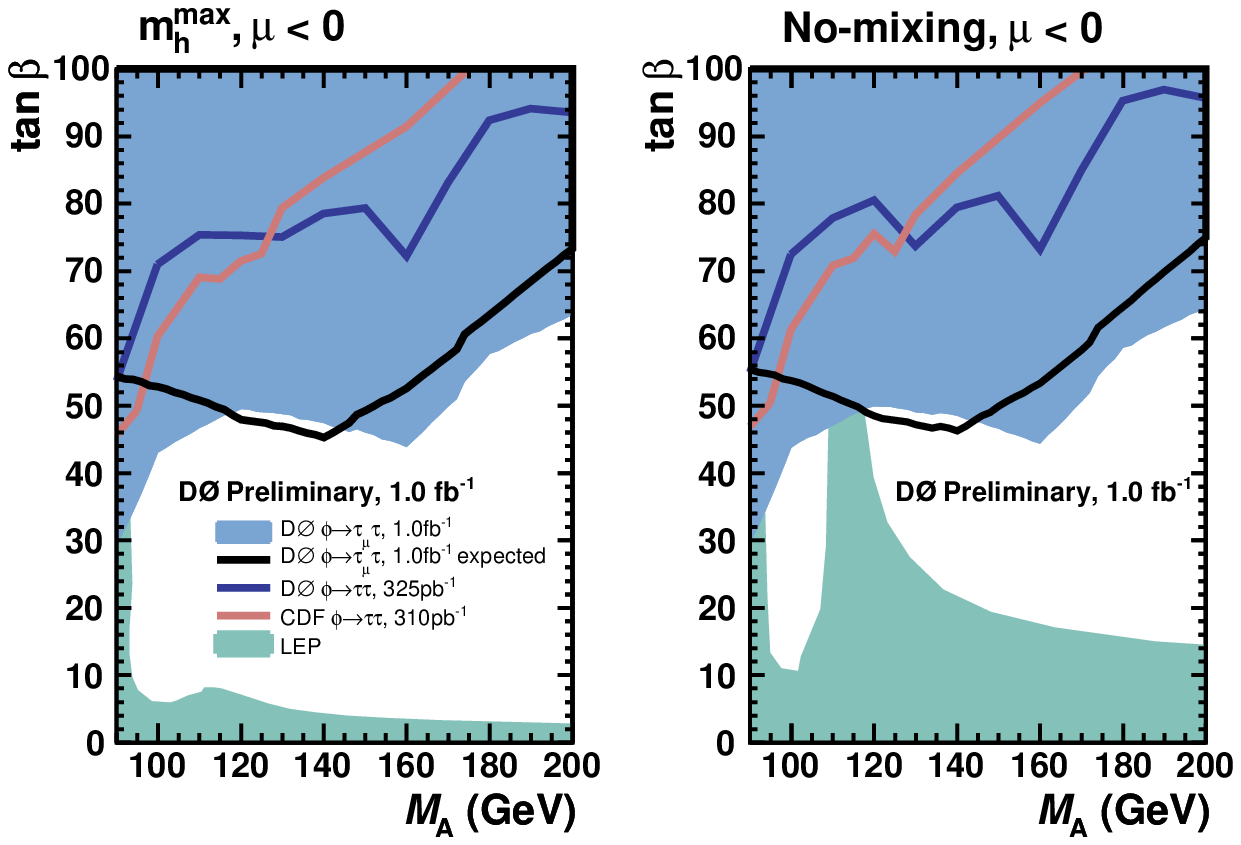}
\includegraphics[width=0.48\textwidth]{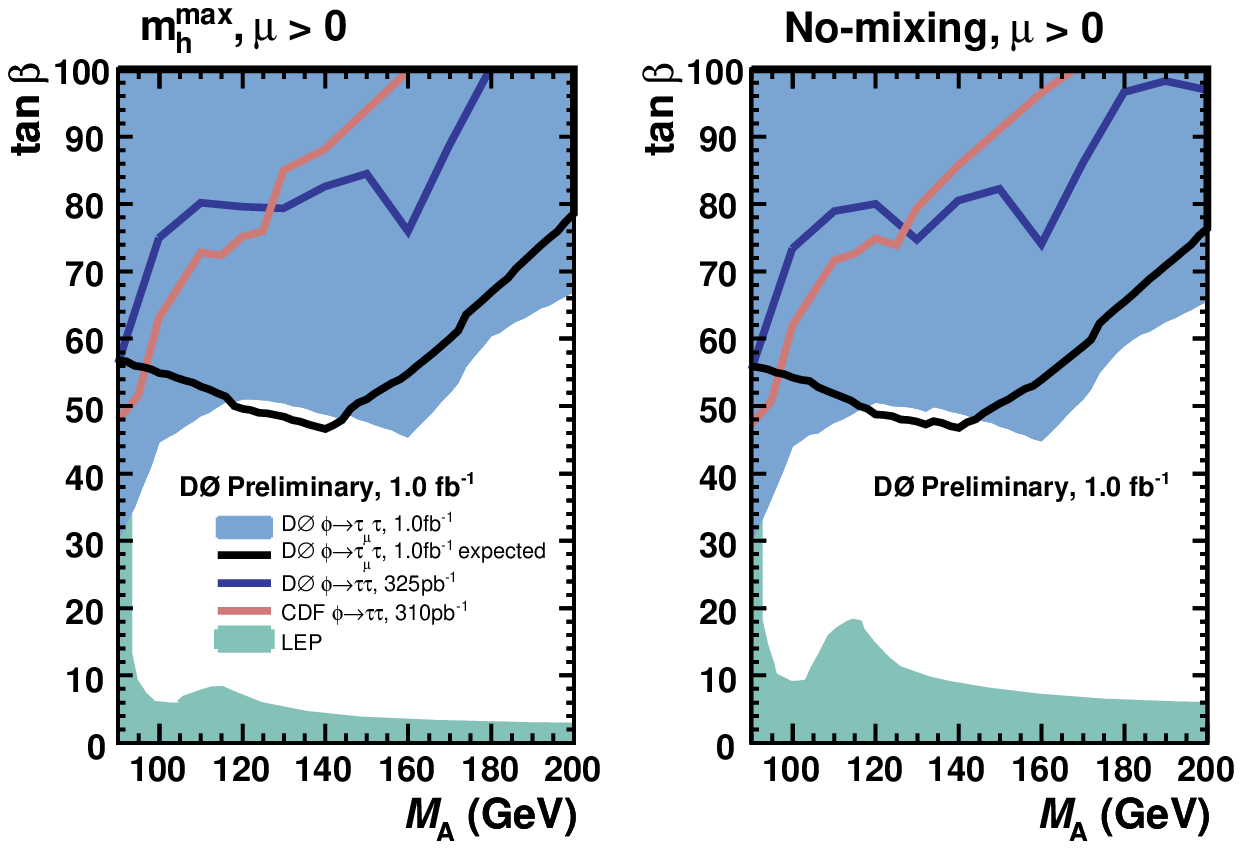}
\caption{Excluded region in the tan$\beta - m_{A}$ plane from D\O\ for a negative (upper) and positive (lower) mass parameter ($\mu$) in the $m^{max}_{h}$ (left) and no-mixing (right) scenarios, along with the LEP limit \cite{lep} and the previous CDF \cite{cdf} and D\O\ \cite{d0} results for $\phi \rightarrow \tau\tau$. These two scenarios are defined by the MSSM parameters in Fig. \ref{htt2}.}
\label{htt1}
\end{center}
\end{figure}

The main background sources in this channel are $Z\rightarrow
\tau^{+}\tau^{-}$ (irreducible), $W +$ jets, $Z\rightarrow
\mu^{+}\mu^{-}/e^{+}e^{-}$ with multi-jet and di-boson events also
contributing. D\O\ has performed a search in the channel where one
of the $\tau$ leptons decays to a $\mu$. The event selection
requires only one isolated muon, separated from the hadronic $\tau$
with opposite sign. The $\tau$ identification is performed with a
neural network. A 20 GeV cut on $M_{W}$, the reconstructed $W$ boson
mass, removes most of the remaining $W$ background. The final
separation of signal from background is achieved with a set of
neural networks, optimized for different Higgs masses and trained on
the visible mass, $m_{vis}$, and $\tau$ and $\mu$ kinematics. The
data are found to be in good agreement with the background-only
expectation. Fig.~\ref{htt1} shows the resulting 95 \% CL exclusion
in the tan$\beta - m_{A}$ plane.

\begin{figure}
\begin{center}
\includegraphics[width=0.48\textwidth]{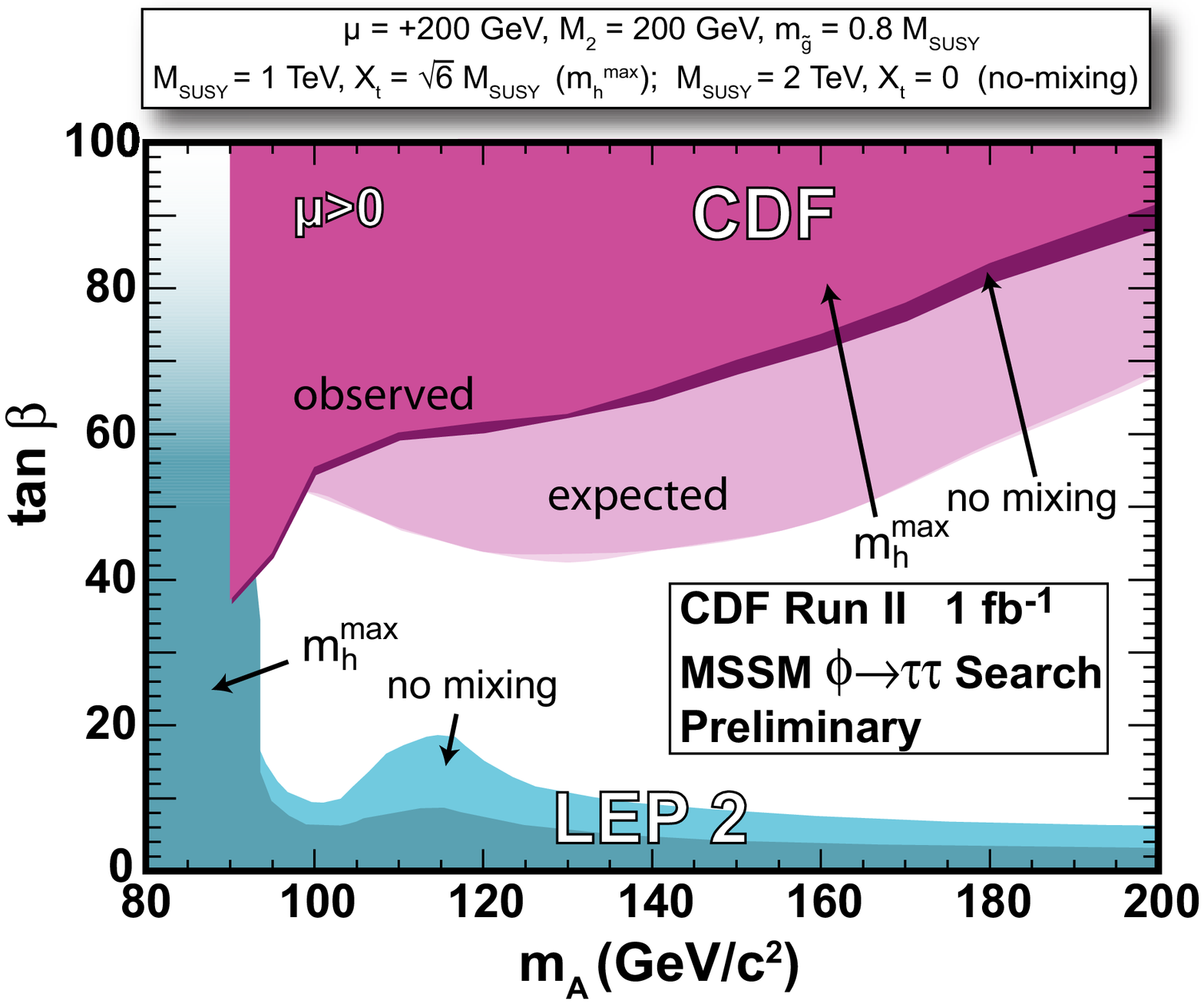}
\includegraphics[width=0.48\textwidth]{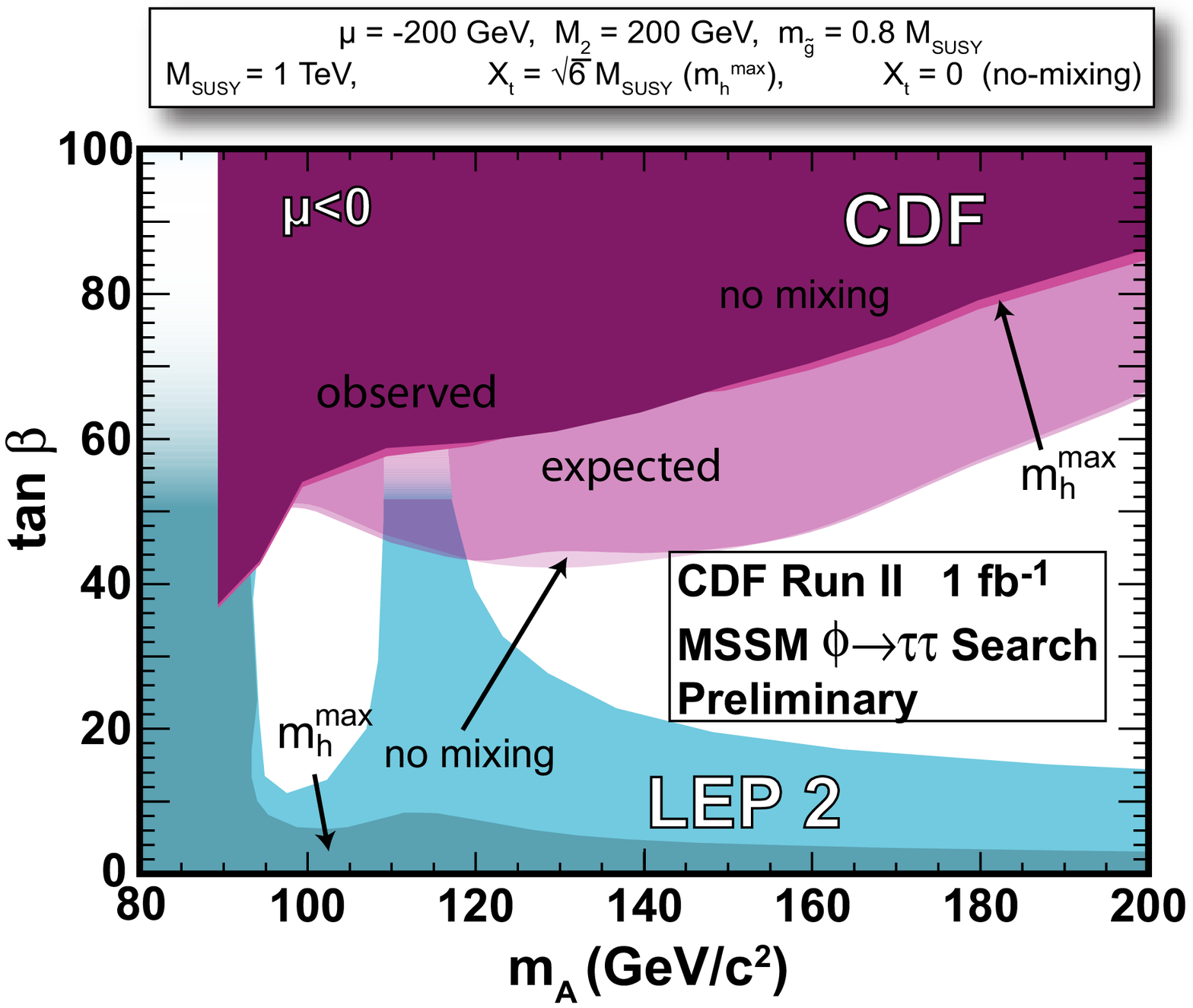}
\caption{\label{htt2}Excluded region in the tan$\beta - m_{A}$ plane from CDF in the $m^{max}_{h}$ and no-mixing scenarios for the cases of a positive (upper) and negative (lower) mass parameter, along with the LEP limit \cite{lep}.}
\end{center}
\end{figure}

CDF has performed a similar search, including channels where one
$\tau$ lepton decays to an electron. The event selection includes an
isolated electron/muon, $\tau$ identification with a variable
cone-size algorithm and jet background suppression with a cut on the
scalar sum of the lepton transverse momentum ($p_{T}$), muon $p_{T}$
and missing transverse energy ($E_{T}$). Most of the $W$ background
is removed by cuts on the relative directions of the visible $\tau$
decay products and the missing $E_{T}$. Limits on cross section
times branching ratio and exclusion regions are derived from the
$m_{vis}$ distribution, the latter is shown in Fig.~\ref{htt2} in
the tan$\beta - m_{A}$ plane. Due to a small excess in the region of
130 GeV $< m_{vis} <$ 160 GeV, the limits are weaker than expected.
However, when all channels ($e\tau,\mu\tau,e\mu$) and possible
search windows are considered the significance of the observed
excess is found to be less than two standard deviations.

\subsection{Higgs $+~b\rightarrow b\bar{b}b$}

D\O\ has carried out a search in this channel using a multi-jet
event sample corresponding to an integrated luminosity of $0.9$
fb$^{-1}$. Candidate events are required to contain at least three
jets with $p_{T} > 15$ GeV, the leading jet must further be above
$40$ GeV and the second jet above $25$ GeV. At least three jets must
be identified as $b$-jets by the standard D\O\ neural network
$b$-tagging algorithm \cite{btagthesis}. A signal is searched for in
the invariant mass spectrum of the two leading $b$-tagged jets. The
simulation of signal and background is performed with PYTHIA
\cite{pythia} or ALPGEN \cite{alpgen} interfaced with PYTHIA and
passed through the detailed detector simulation. The dominant
background is multi-jet production and is estimated from the data
outside the signal search region. The signal acceptance is found to
be 1.7-2.6\% depending on the Higgs mass. As no significant excess
is observed, limits are set. Cross sections down to 20 pb are
excluded for Higgs masses up to 170 GeV.

\subsection{Higgs $+~b\rightarrow \tau^{+}\tau^{-}b$}

A single muon event sample collected by D\O, corresponding to an
integrated luminosity of $0.3$ fb$^{-1}$, is used to search for the
final state where one $\tau$ decays hadronically and the other to a
$\mu$. Candidate events are required to have one $\mu$ with a $p_{T}
> $ 12~GeV, a hadronic $\tau$ with an opposite sign to the $\mu$, which is identified by using a neural
network, and at least one $b$-jet with $p_{T} > 15$ GeV, identified
using an impact parameter b-tagging tool. The three major
backgrounds are QCD multi-jet production, $Z+jets \rightarrow
\mu\tau + jets$ and $t\bar{t}\rightarrow b\bar{b}\mu\tau$. The QCD
multi-jet and Z+jets backgrounds are estimated from data and the
other backgrounds are simulated using ALPGEN interfaced with PYTHIA.
The signal is simulated using PYTHIA. After $b$-tagging the largest
contribution is from $t\bar{t}$ events, which are removed using a
neural network based upon kinematic variables. In the absence of any
excess limits are set using the invariant mass distribution,
calculated from the 4-vectors of the $\mu$, hadronic $\tau$ and
missing $E_{T}$. Fig.~\ref{htt1} shows the resulting 95 \% CL
exclusion in the tan$\beta - m_{A}$ plane.

\begin{figure}[h]
\begin{center}
\includegraphics[width=0.48\textwidth]{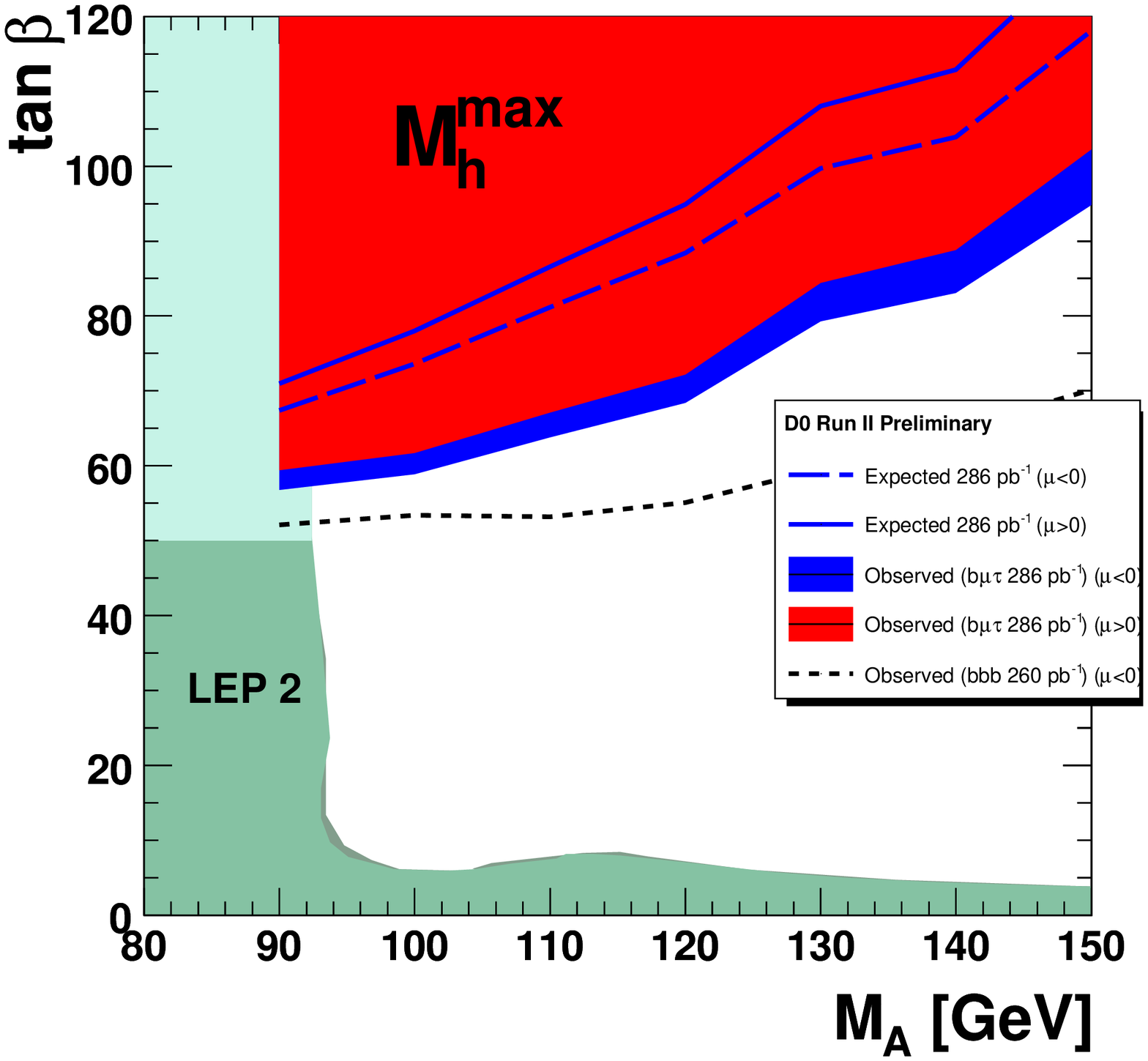}
\includegraphics[width=0.48\textwidth]{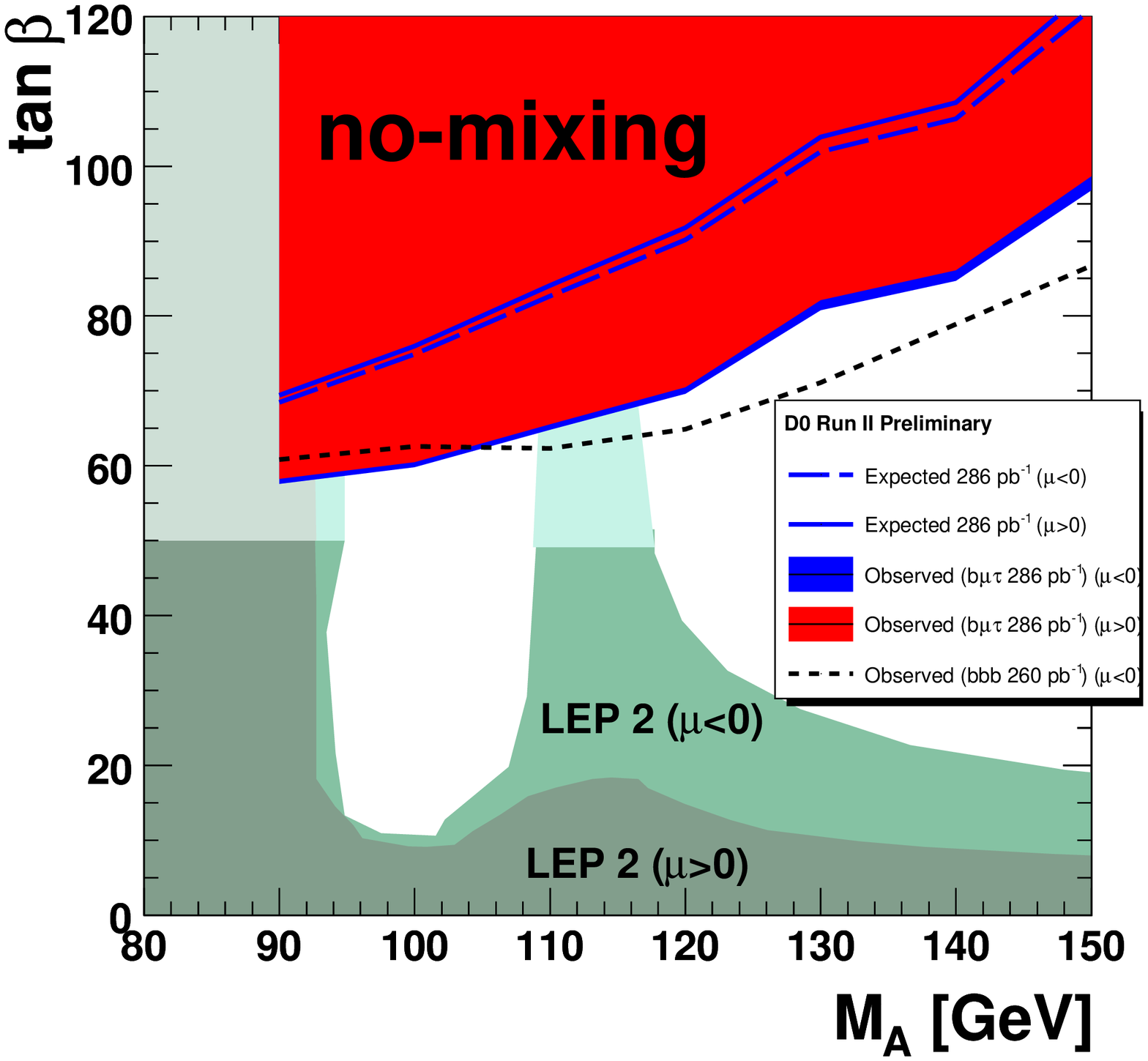}
\caption{Excluded region in the tan$\beta - m_{A}$ plane in the
$m^{max}_{h}$ (upper) and no-mixing (lower) scenarios along with the
LEP limit \cite{lep} and previous D\O\ \cite{d0hb} limits for $\phi
b\rightarrow b\bar{b}b$. The two scenarios are defined by the MSSM
parameters in of Fig. \ref{htt2}.}
\end{center}
\end{figure}

\section{Limits on non-SM Higgs $\rightarrow \gamma \gamma$}

Though the Higgs to photon branching ratio is negligible in the SM,
some extensions predict a significantly larger value. A fermiophobic
Higgs does not couple to fermions at all and a Top-color Higgs has
zero coupling to all fermions except the top quark. Such models
would hence result in an enhanced rate of Higgs bosons decaying to
photons.

D\O\ has searched for Higgs bosons in $3\gamma + X$ final states in
data corresponding to an integrated luminosity of $0.8$ fb$^{-1}$.
The event selection includes three isolated photons with $E_{T} >
15$ GeV within $|\eta| < 1.1$ (central calorimeter). The combined
transverse momentum of the three photons is further required to be
larger than 25 GeV. 0 events are selected with a total expected
background of $1.1 \pm 0.2$ events. The background is dominated by
direct triple photon production with a small contribution from QCD
and $Z/W+X$ processes. No excess is observed and hence excluded
fermiophobic Higgs masses are calculated. This search excludes a
fermiophobic Higgs below 80 GeV for a charged Higgs mass below 100
GeV and tan$\beta = 30$.

\section{Conclusions}

The preliminary results presented at this conference by the CDF and
D\O\ collaborations, together with the recent performance of the
experiments and the Tevatron, are very encouraging for the Higgs
searches at Run II. The 1 fb$^{-1}$ searches for Higgs bosons beyond
the SM, in the MSSM scenario and other extensions, show very
promising sensitivity and have already produced new powerful limits
on $h/H/A \rightarrow \tau\tau/b\bar{b}$ and $h\rightarrow
\gamma\gamma$. New MSSM results can be expected from both
experiments shortly, with both more data and improvements to the
analyses themselves. Work will also focus on combining the
complimentary results from the different channels and from both
experiments.

Having successfully accomplished analyses of the first fb$^{-1}$ of
Run II data, CDF and D\O\ are confidently looking forward to
exploring the almost $3$ fb$^{-1}$ of data per experiment which has
already been written to tape, and the $\sim$8~fb$^{-1}$\ total per
experiment expected by the end of Run II.


\begin{thebibliography}{999}
\bibitem{mssm} Dimopoulos S and Georgi H 1981 {\it Nucl. Phys.} B {\bf 193}, (1981)
150.
\bibitem{topcolor} Hill C T, {\it Phys. Lett.} B {\bf 266} (1991) 419.
\bibitem{haber} Haber H E, Kane G L and Sterling T, {\it Nucl. Phys.} B {\bf
161} (1979) 493.
\bibitem{cdf_web_page} http://www-cdf.fnal.gov
\bibitem{dzero_web_page} http://www-d0.fnal.gov
\bibitem{lep} Schael S {\it et al.}, {\it Eur. Phys. J.} C {\bf 47}
(2006) 547-587.
\bibitem{cdf} Abulencia A {\it et al.}, {\it Phys. Rev. Lett.} {\bf
96} (2006) 011802.
\bibitem{d0} Abazov V M {\it et al.}, {\it Phys Rev. Lett.} {\bf 97}
(2006) 121802.
\bibitem{btagthesis} T. Scanlon, ``b-Tagging and the Search for
Neutral Supersymmetric Higgs Bosons at D\O\/'',
FERMILAB-THESIS-2006-43.
\bibitem{pythia} Sj\"ostrand T, L\"onnblad L, Mrenna S and Skands P, PYTHIA 6.3 Physics and Manual {\it Preprint}
hep-ph/0308153 (2003).
\bibitem{alpgen} Mangano M L, Moretti M, Piccinini F, Pittau R and Polosa A D, {\it J. High Energy Phys.} {\bf
0307} (2003) 001.
\bibitem{d0hb} Abazov V M {\it et al.}, {\it Phys Rev. Lett.} {\bf 95} (2005) 151801.
\end{thebibliography}
\end{document}